\documentclass[aps,prl,reprint,groupedaddress]{revtex4-1}

\usepackage{ae,amsmath,amssymb,amsfonts,bbm,amsthm}
\usepackage[english]{babel}
\usepackage{braket}
\usepackage{graphicx}
\usepackage{calc}
\usepackage{color}
\usepackage{transparent}
\usepackage{dcolumn}
\usepackage{hyperref}
\usepackage{here}

\usepackage{adjustbox}

\usepackage{units}
\newcommand{\Mod}[1]{\ (\text{mod}\ #1)}
\begin{document}

\title{Ultrafast Long-Distance Quantum Communication with Static Linear Optics}

\author{Fabian Ewert}
\email[]{ewertf@uni-mainz.de}
\affiliation{Institute of Physics, Johannes Gutenberg-Universität Mainz, Staudingerweg 7, 55128 Mainz, Germany}
\author{Marcel Bergmann}
\email[]{marcel.bergmann@uni-mainz.de}
\affiliation{Institute of Physics, Johannes Gutenberg-Universität Mainz, Staudingerweg 7, 55128 Mainz, Germany}
\author{Peter van Loock}
\email[]{loock@uni-mainz.de}
\affiliation{Institute of Physics, Johannes Gutenberg-Universität Mainz, Staudingerweg 7, 55128 Mainz, Germany}

\begin{abstract}
We propose a projection measurement onto encoded Bell states with a static network of linear optical elements. By increasing the size of the quantum error correction code, both Bell measurement efficiency and photon-loss tolerance can be made arbitrarily high at the same time. As a main application, we show that all-optical quantum communication over large distances with communication rates similar to those of classical communication is possible solely based on local state teleportations using optical sources of encoded Bell states, fixed arrays of beam splitters, and photon detectors. As another application, generalizing state teleportation to gate teleportation for quantum computation, we find that in order to achieve universality the intrinsic loss tolerance must be sacrificed and a minimal amount of feedforward has to be added.
\end{abstract}
\maketitle

\paragraph{Introduction.\label{sec:motivation}}

Since the ground breaking work of Duan et al. (DLCZ, \cite{DLCZ}) who showed that long-distance quantum communication (LDQC) is possible with linear optics and atomic-ensemble quantum memories, numerous advanced versions of their quantum repeater protocol have been proposed \cite{SangouardRMP}. However, the probabilistic nature of entanglement distribution over lossy channels, purification, and swapping makes this type of nested quantum repeaters extremely slow, relying on two-way classical communication and long-lived quantum memories.

In recent years, various proposals have been made to employ quantum error correction (QEC) codes for LDQC. Since these codes suppress errors deterministically, long waiting times and two-way classical communication (and hence the use of quantum memories) can be, in principle, completely avoided. While one class of schemes focused on the correction of operational errors \cite{Jiang09,Munro10,Bernardes12,Bratzik14}, another class did include QEC against transmission losses making high-rate loss-tolerant \cite{Munro12,ATL,pant2016rate} or even fully fault-tolerant \cite{Fowler10,LLPRL,namiki16,muralidharan2016} LDQC possible. These latter schemes are limited only by the speed of the local gate operations and thus, they approach rates as obtainable in classical communication. Our scheme also allows for ultrafast LDQC, but unlike \cite{Munro12,Fowler10,LLPRL} it does so in an all-optical fashion without the use of difficult local quantum gates (implementable via local nonlinear matter-light interactions \cite{Munro12,LLPRL}).

\begin{figure}[b]
	\includegraphics[scale=0.9]{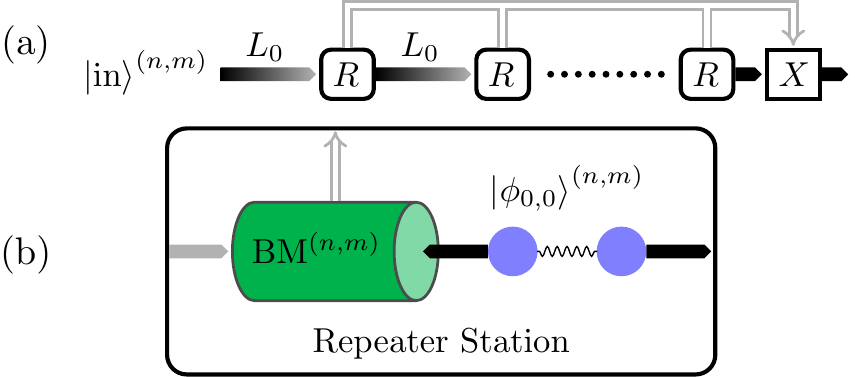}
	\caption{One-way communication scheme: (a) To send a quantum state $\ket{\rm in}^{(n,m)}$ over a long distance, repeater stations ($R$) at shorter distances $L_0$ are used to recover the qubit from accumulated losses (fading arrows). A classical signal (double line) defines a single Pauli correction $X$ at the receiver. (b) Each repeater station consists of an encoded Bell state and a highly efficient, loss-resistant, logical Bell Measurement [BM$^{(n,m)}$] acting on the incoming signal and one half of the Bell state. The other half of the Bell state is sent to the next station along with the result of the BM (classical signal). \label{fig:comm_scheme}}
\end{figure}

For this purpose, by employing a certain version of loss-tolerant parity codes \cite{RalphGilchrist05,Munro12,LLPRL}, we suggest sending encoded qubit states directly, which are then subject to a Bell measurement (BM) together with locally prepared, encoded Bell states after every few kilometers (see Fig.~\ref{fig:comm_scheme}). These local state teleportations allow for a nondestructive loss-error syndrome detection and a qubit state recovery in one step. The use of QEC by teleportation \cite{Knill} along the channel is conceptually similar to the protocol of Ref.~\cite{LLPRL}. However, in our scheme, every teleportation is performed with optical (encoded) Bell states and linear optical elements
\footnote{In terms of resources, our all-optical scheme is actually most close to that of Ref.~\cite{ATL} which employs nonlocally distributed loss-resistent cluster states and standard (nonlogical) linear-optics BMs for entanglement connection. However, there, in order to suppress the effect of losses, fast feedforward operations are required at every repeater station to separate successful from failed BM events (depending only on the local BM results and independent of the classical information about neighboring BM outcomes like in multiplexed standard repeaters \cite{Collins}).}.
It turns out that the encoding has two positive effects: the larger the code is, the more efficient the ideal BM (despite the linear-optics constraint \cite{CalsamigliaNL}) {\it and} the higher the amount of tolerable photon loss becomes. In contrast to the all-optical scheme of Ref.~\cite{ATL}, our logical BMs are conceptually different and work entirely without feedforward. This not only reduces the local operation times, but also makes on-chip integration along an optical fiber channel more feasible, as optical switching in this case is very sensitive to loss \cite{QTonChip14,LossMultiplObrien14}
\footnote{Another important complication when integrating feedforward on a chip is the need for synchronizing the optical and electrical signals. To be compatible with optical pulses much shorter than \unit[1]{ns}, fast integrated electronic circuits (including measurements) with electronic bandwidths much greater than \unit[1]{GHz} are required. Therefore, instead of real-time feedforward, simple postselection is often employed \cite{QTonChip14}.}.
In an extended version of this work \cite{Ewert2016draft}, we give further details on the loss resistance of our scheme and we show that it is also robust against a variety of additional errors such as depolarizing errors and detector inefficiencies (loss and dark counts) by performing a detailed secure-key-rate analysis. It is also demonstrated there that the scheme still works when photon-number-resolving detectors (as considered here) are replaced by on-off detectors. Beyond quantum communication, here we show that for universal quantum computation, our encoded BM ceases to work under full loss tolerance, but ideal scalable quantum computation with linear optics is still possible with some but less feedforward compared to the Knill, Laflamme, and Milburn (KLM, \cite{KLM}) and cluster-based \cite{RaussendorfBriegel,Nielsen,BrowneRudolph} approaches.

\paragraph{Encoded Bell measurements.\label{sec:encodedBM}}

The quantum parity code [QPC($n,m$)] \cite{RalphGilchrist05,LLPRL} encodes a logical qubit into $n m$ physical qubits. The code can be understood as having three different levels of encoding. On the lowest level, which we call the physical level, we have standard dual-rail (DR, two-mode) qubits. These are typically realized by two orthogonal polarization modes of photons $\{\ket{0} = \ket{H},\ket{1} = \ket{V}\}$, but also other realizations like spatial or temporal modes are possible. On the second level of encoding, the block level, $m$ physical qubits are collected to represent a block qubit $\{\ket{0}^{(m)} = \ket{H}^{\otimes m}, \ket{1}^{(m)} = \ket{V}^{\otimes m}\}$. This repetition part of the code is crucial for the loss robustness as we see later. The highest encoding level is the logical level. Here $n$ block qubits are used to construct the logical qubits as $\ket{\pm}^{(n,m)} = \left[\ket{0}^{(m)} \pm \ket{1}^{(m)} \right]^{\otimes n} / \sqrt{2^n} = [\ket{\pm}^{(m)}]^{\otimes n}$. The codewords are then naturally obtained by $\ket{\pm}^* = \left[\ket{0}^* \pm \ket{1}^* \right]/ \sqrt{2}$, where the $*$ denotes the encoding level [blank for physical, $(m)$ for block and $(n,m)$ for logical]. 

\begin{figure}[t]
	\includegraphics[scale=1]{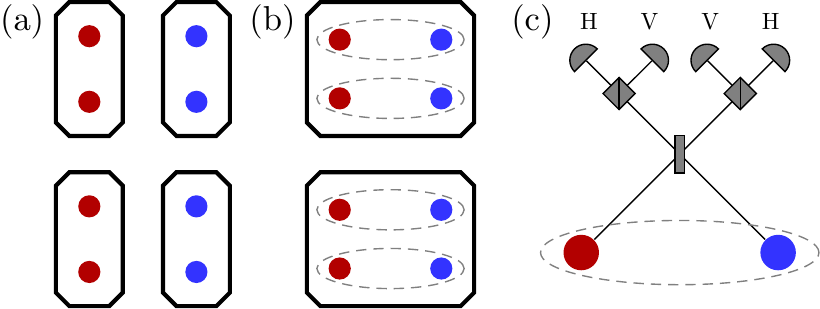}
	\caption{Block structure and Bell measurement: (a) The block structure for two QPC(2,2)-qubits. The polarization qubits on the left (red) belong to the incoming signal and are thus subject to channel errors, while those on the right (blue) are part of the encoded Bell state provided in the repeater station. (b) In a Bell state in QPC encoding the qubits are joined blockwise.  The dashed ellipses highlight physical-level qubit pairs that are combined at the BM. (c) Optical BM setup on the physical level adapted to polarization encoding.\label{fig:BM_setup}}
\end{figure}

In all three encoding levels the four Bell states are defined as
\begin{align}
	\ket{\phi_{k,l}}^* = \tfrac{1}{\sqrt{2}}\left[\ket{0,k}^* + (-1)^l \ket{1,1-k}^*\right],  \label{eq:bell_states}
\end{align}
with $k,l\in\{0,1\}$. A Bell measurement has to distinguish between these four Bell states. On the physical level, this can be partially achieved by combining the two polarization qubits at a $50:50$ beam splitter followed by polarizing beam splitters and photon detectors [see Fig.~\hyperref[fig:BM_setup]{\ref*{fig:BM_setup}(c)}]. Unique click patterns are obtained for $\ket{\phi_{1,0}}$ and $\ket{\phi_{1,1}}$, whereas the states $\ket{\phi_{0,l}}$ are indistinguishable from each other. Thus, the overall BM efficiency is $50\%$, which is optimal for dual-rail encoding without ancilla photons or feedforward \cite{CalsamigliaNL}.

Our approach to a BM on QPC($n,m$)-encoded qubits is based on the observation that Bell states of the higher encoding levels can be represented in terms of lower-encoding-level Bell states.
\begin{align}
	\ket{\phi_{k,l}}^{(m)} &\cong \frac{1}{\sqrt{2^{m-1}}}\sum_{\vec{r} \in A_{l,m}} \bigotimes_{i=1}^m \ket{\phi_{k,r_i}},\label{eq:phys2block}\\
	\ket{\phi_{k,l}}^{(n,m)} &\cong \frac{1}{\sqrt{2^{n-1}}}\sum_{\vec{s} \in A_{k,n}} \bigotimes_{i=1}^n \ket{\phi_{s_i,l}}^{(m)},\label{eq:block2logic}
\end{align}
where the index set is defined as $A_{l,m} = \Set{\vec{r}\in\{0,1\}^m | \sum_{i=1}^m r_i = l \Mod 2 }$
\footnote{A derivation of these formulas can be found in the Supplemental Material.}.
These relationships between Bell states of different encoding levels show that a logical BM can be realized by $n m$ simultaneous standard BMs on the physical level.

Note that the above representations \eqref{eq:phys2block},\eqref{eq:block2logic} only hold after an appropriate reordering of the modes (indicated by $\cong$). Quite naturally, the photons of two logical qubits in a Bell state are each paired with their equivalent [see Fig.~\hyperref[fig:BM_setup]{\ref*{fig:BM_setup}(a)} and \hyperref[fig:BM_setup]{\ref*{fig:BM_setup}(b)}]. In the following this reordering is omitted in the notation.

A Bell measurement on the block level is limited by the same $\tfrac{1}{2}$-efficiency as a physical BM, because the index $k$ determining whether a BM on the physical level is successful is the same for all physical Bell states within a block Bell state. On the other hand, the index $l$ is always identified correctly for $k=1$, because in that case the values $r_i$ are all accessible. On the logical level, the situation is quite different. The index $k$ is always identified correctly, because the values $s_i$ from the block level are always available. Additionally, almost every time the index $l$ will now be identified correctly as well, since it suffices to identify it in a single block. The only case where this is not possible is when all block-level Bell states are $\ket{\phi_{0,l}}^{(m)}$ states. This can only occur in the states $\ket{\phi_{0,l}}^{(n,m)}$ with a statistical weight of $2^{1-n}$. Consequently, the chance to identify a logical Bell state correctly, i.e. the logical BM efficiency, is $1-2^{-n}$.

In addition to boosting the BM efficiency to near unity, the QPCs also protect the Bell measurement against photon loss. In accordance with our communication scheme depicted in Fig.~\ref{fig:comm_scheme}, we assume that only the photons of one logical qubit participating in the BM are affected by loss
\footnote{A discussion on the effect of imperfect ancillary Bell states including loss on these is presented in the Supplemental Material.}.
Furthermore, we make the usual assumption that the probability to lose a photon ($1-\eta$) is the same for all modes of a logical qubit. The probability of a successful logical Bell measurement in the presence of loss quantified by $\eta$ can be derived from the Bell state representations \eqref{eq:phys2block} and \eqref{eq:block2logic}. To identify the value of $k$ in the state $\ket{\phi_{k,l}}^{(n,m)}$, a correct identification of every value $s_i$ is required; i.e., in every block the first index must be determined. Equation~\eqref{eq:phys2block} shows that this is possible, as long as in every block at least one physical Bell measurement identifies the first index. Since this is guaranteed as long as in every block at least one of the $m$ photons (belonging to that logical qubit subject to loss) is not lost, the probability of correctly identifying $k$ is given by $\left[1-\left(1-\eta\right)^m\right]^n$. In addition, in order to identify the index $l$ in $\ket{\phi_{k,l}}^{(n,m)}$, it must be determined in at least one block. The probability to identify $l$ in a block is given by $\tfrac{\eta^m}{2}$, because all values $r_i$ are required, which means there are no photons lost at all, and because only the states $\ket{\phi_{1,l}}^{(m)}$ allow us to detect $l$ with standard linear optical means. In other words, for the relevant logical qubit subject to loss, at least one photon must be left in every block and at least one block must remain entirely uncorrupted. The success probability of the logical BM is therefore given by
\begin{align}
	p = \left[1-\left(1-\eta\right)^m\right]^n - \left[1-\left(1-\eta\right)^m - \tfrac{\eta^m}{2}\right]^n, \label{eq:psucc}
\end{align}
where the second term expresses that all terms where enough photons were left to identify $k$ but no block allowed to identify $l$ have to be discarded
\footnote{An alternative, maybe more intuitive way to derive this probability is given in the Supplemental Material.}.

\begin{table}[t]
\caption{BM success probability $p$ in $\%$ for various QPC($n,m$) and varying loss $\eta$.\label{table:probs}}
\begin{ruledtabular}
\begin{tabular}{rddddddd}
($n,m$) & \multicolumn{1}{r}{$\eta=1$} & 0.99 & 0.95 & 0.90 & 0.75 & 0.50 & 0.30 \\
\hline
 (1,1) &  50 & 49.5 & 47.5 & 45 & 37.5 & 25 & 15 \\
 (2,2) & 75 & 73.99 & 69.66 & 63.79 & 44.82 & 17.19 & 4.39 \\
 (3,10) & 87.5 & 83.56 & 65.61 & 43.71 & 8.21 & 0.15 & 0.00 \\
 (6,5) & 98.44 & 97.92 & 94.69 & 87.74 & 52.86 & 7.68 & 0.29 \\
 (10,3) & 99.90 & 99.87 & 99.51 & 97.95 & 77.77 & 13.77 & 0.28 \\
 (23,5) & 100.00 & 100.00 & 100.00 & 99.95 & 92.44 & 15.03 & 0.05
\end{tabular}
\end{ruledtabular}
\end{table}

Table~\ref{table:probs} shows the attainable BM efficiency for various amounts of loss.  It indicates that the QPCs indeed protect the logical qubit from loss as long as $\eta>0.5$, and also that, in general, $n$ should be chosen sufficiently larger than $m$: a too large $m$ increases the chance of corrupting every block. However, a too small $m$ risks corrupting all photon pairs in a block. Increasing $n$ on the other hand only gives more blocks, thus increasing the chance to get at least one without any corruption. Conceptually, this is the most important result obtained here: a larger code with a larger number of blocks $n$ results in a higher linear-optics BM efficiency and a higher loss tolerance at the same time. This is different from other BM schemes where the loss tolerance is decreasing which has to be counteracted by additional quantum error correction \cite{Lee} or fast feedforward operations \cite{ATL}.

\paragraph{Long-distance quantum communication.\label{sec:ldqc}}

\begin{figure}[t]
\centering
\includegraphics[width=0.475\textwidth]{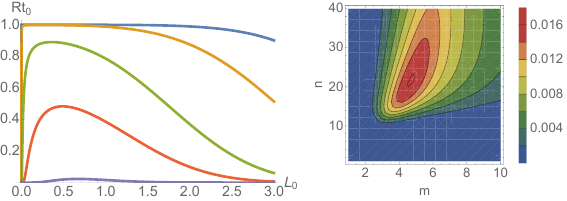}
\caption{Left: Total success probability $R t_0$ vs repeater spacing $L_0$ in km for a communication distance of $L=\unit[1\,000]{km}$ and various encodings [from bottom to top, (10,3),(13,4),(16,4),(23,5),(35,6)]. Right: Inverse of the cost function $C_{\unit[1\,000]{km}}$ as a function of the code parameters $n,m$. At every point the optimal repeater spacing $L_0$ is chosen. The most cost-efficient code is $(23,5)$ with a repeater spacing of $L_0\approx \unit[2.36]{km}$ yielding $R t_0 = 77.62\%$.}\label{fig:plotspsucc}
\end{figure}

To send a QPC($n,m$)-encoded qubit state over a total distance $L$ we propose placing repeater stations after every channel segment with length $L_0$. At every station an encoded Bell state $\ket{\phi_{0,0}}^{(n,m)}$ is available on demand (i.e., created and consumed locally), and a logical BM is performed on one half of the Bell state together with the incoming encoded qubit (see Fig.~\ref{fig:comm_scheme}). Between two stations every physical qubit suffers from loss according to a transmission coefficient of $\eta=\exp(-L_0/L_{\rm att})$ (with attenuation length $L_{\rm att} = \unit[22]{km}$). Whenever the BM succeeds, the qubit state is recovered from loss and appears at the other half of the Bell state to be sent to the next station \cite{Knill,Bennett}. The total success rate
\footnote{Since photon loss alone is incapable of inducing bit-flip or phase-flip errors on dual-rail qubits, the total success rate is also the secure key rate of our scheme, because the quantum bit error rate vanishes.}
of the communication scheme is then given as $R=p^\frac{L}{L_0}/t_0$, where $t_0$ is the elementary time needed at every repeater station until the incoming signal qubit has been processed and a fresh encoded Bell state is ready for teleporting and error correcting the next qubit.

In addition to the repeater success probability $R t_0$, which is depicted in Fig.~\ref{fig:plotspsucc} as a function of the repeater spacing $L_0$ for various code sizes, we are also interested in the cost effectiveness of our communication scheme. To this end, we define the cost function $C_L=\tfrac{n m}{R t_0 L_0}$ for a given total distance $L$ similar to that in \cite{LLPRL}. It relies on the assumption that the cost for creating the ancillary encoded Bell states at every repeater station scales linearly in $n m$ (an all-optical method for state generation based on coherent photon conversion \cite{langford11} that achieves this kind of scaling is presented in the Supplemental Material). The inverse $1/C_L$, which corresponds to the repeater success probability per photons used, is also shown in Fig.~\ref{fig:plotspsucc} for a total communication distance $L=\unit[1\,000]{km}$. Figure~\ref{fig:plotspsucc} indicates that total success rates extremely close to $R=1/t_0$ can be achieved even for fairly large repeater spacings, but in terms of cost effectiveness a rate of about $R \approx 0.75/t_0$ yields better results
\footnote{The recent benchmarks for repeaterless communication, the so-called Takeoka-Guha-Wilde (TGW) bound \cite{TGW} and its refinements \cite{PLOB} relating to the transmission rate per mode, are comfortably beaten by our scheme; see \cite{Ewert2016draft}.}.

Furthermore, for comparison the cost function $C_L$ can also be applied to the case of (near) perfect Bell measurements on the physical level \cite{Zaidi,Grice,Ewert} (e.g., realized with additional atomic processing qubits \cite{LLPRL}). While these better BMs allow for an efficient use of smaller codes, we found that for the optimal choices of $n$, $m$, and $L_0$ the ratio $C_L(p_{BM}=0.5)/C_L(p_{BM}=1) \approx 3$ is almost independent of the communication distance $L$. This imposes a limit on how much more expensive perfect BMs should be compared to a standard optical BM with efficiency $\tfrac{1}{2}$.

We should also consider the effect of the elementary processing time $t_0$. Since our logical Bell states are assumed to be available on demand, $t_0$ corresponds only to the duration of the linear-optics processing with photon detection. Compared to those times required in a matter-based scheme with $t_0 \sim 1\mu s$ (even assuming future enhanced ion-cavity coupling strengths \cite{LLPRL}) or an all-optical scheme including feedforward \cite{ATL} with $t_0 \sim 10 ns$ (provided all circuits can be integrated \cite{Prevedel}), corresponding to rates $R\sim \unit[]{MHz}$ or $R\sim \unit[0.1]{GHz}$, respectively, our static linear optical scheme allows, in principle, for $\unit[]{GHz}-$rates and beyond
\footnote{Provided that the signal repetition rates and the detector bandwidths are sufficiently high.}.
For our scheme to be free of feedforward at the intermediate stations, the updated (logical) Pauli frame after each teleportation must be classically communicated to the end of the channel for a final Pauli correction
\footnote{Any active operations at a finite gate speed would, of course, reduce the achievable rates despite the complete avoidance of such operations along the channel. However, for certain applications such as quantum key distribution, a final Pauli frame correction is not needed. More generally, often simple postselection suffices.}.
In general, let us discuss next universal gates and gate teleportation based on our encoded BM.

\paragraph{Quantum gate teleportation and quantum computation.}\label{sec:computation}

The physical Pauli operators of the QPC$(n,m)$ may be denoted as $X_{i,j},Y_{i,j},Z_{i,j}$, with $i=1...n$ and $j=1...m$ labeling the $(i,j)$th DR qubit, while the logical operators are $X^{(n,m)}=X_{i,1}...X_{i,m}$ (for any $i$) and $Z^{(n,m)}=Z_{1,j}...Z_{n,j}$ (for any $j$) \cite{LLPRL}. Therefore, Pauli logic can be performed directly via suitable Pauli gates on the DR qubits. This is sufficient for the final Pauli frame correction in our LDQC scheme as well as for quantum key distribution applications. More generally, logical $X$ and $Z$ rotations are then given by $\exp[-iX^{(n,m)}\theta/2]$ and $\exp[-iZ^{(n,m)}\theta/2]$, respectively, and for any $\theta\notin \pi\mathbb{Z}$ and $n>1,m>1$, an entangling operation is needed that acts on the physical qubits. Based on our encoded linear-optics BM, we can use logical gate teleportation with suitable encoded offline resource states \cite{Chuang} to implement arbitrary Clifford computations [including logical two-qubit gates such as \textsc{cnot}$^{(n,m)}$] in an intrinsically loss-tolerant fashion with no need for feedforward between the Clifford gates (and with only a final Pauli frame correction). This is a huge simplification compared to KLM \cite{KLM} who require feedforward for every single \textsc{cnot} and additional QEC codes to correct photon-loss errors. However, for universality, any single-qubit gate of KLM can be performed directly on the DR qubits, whereas in our general QPC$(n,m)$ scheme, the logical non-Clifford gates do not allow for a static BM-based gate teleportation or a nonentangling transversal gate application. Therefore, for universality, we have to sacrifice the intrinsic loss tolerance and employ the most simple versions of the QPC such as QPC$(n,1)$ 
\footnote{If feedforward is allowed, there are ways to achieve universality in a loss-tolerant or even fault-tolerant fashion on QPC-type encoding \cite{Hayes1,Hayes2}.}.
In this case, an arbitrary logical $X$ rotation $\exp[-iX^{(n,1)}\theta/2]$ can be done via the same rotation $\exp[-iX_{i,1}\theta/2]$ directly on the $i$th DR qubit (for any $i$) and the remaining set of Clifford operations [including a single-qubit $\pi/2$-rotation $\exp[-iZ^{(n,1)}\pi/4]$ for universality] can be achieved through gate teleportation using the static linear-optics BM scheme. Since QPC$(n,1)$ is enough to realize arbitrarily efficient BMs (for sufficiently high $n$), efficient linear-optics quantum computation is possible provided a little, simple Pauli feedforward is added every time when a sequence of Clifford gates is followed by a non-Clifford gate. In terms of feedforward, this is also a simplification compared to existing schemes
\footnote{Another huge simplification compared to KLM is that in this loss-free case on-off detectors are sufficient.},
where every two-qubit gate requires Pauli corrections on randomly selected physical qubits (for KLM \cite{KLM}) or even non-Pauli feedforward is needed (for one-way quantum computation \cite{RaussendorfBriegel,Prevedel}).

\paragraph{Discussion and conclusions.}

We proposed an efficient linear-optics BM onto QPC-encoded Bell states and showed that, by incorporating protection against transmission losses, it can be used to realize ultrafast high-rate LDQC in an all-optical fashion. With no need for matter qubits (neither as quantum memories nor as local quantum processors) or feedforward operations, our communication scheme is most suitable to be integrated along an optical fiber channel via chips that contain quantum sources \cite{Silverstone14,Silverstone2015,Spring13}, interferometers \cite{Peruzzo11}, and photon detectors \cite{Calkins13}. Encoded-state preparations may be based either on nonlinear optical techniques \cite{langford11} or on linear optics \cite{ATL,pant2016rate,Ewert2016draft}, then including feedforward 
\footnote{For details on the state generation schemes see the supplemental material.}.
\begin{acknowledgments}
We acknowledge support from Q.com (BMBF) and Hipercom (ERA-NET CHISTERA).
\end{acknowledgments}

\bibliography{repeaterbib}
\onecolumngrid

\section*{Supplemental material}
\setcounter{table}{0}
\setcounter{equation}{0}
\setcounter{lem}{0}
\setcounter{Theorem}{0}
\renewcommand{\theequation}{S\arabic{equation}}
\renewcommand{\thetable}{\Roman{table}s}

\subsection{Bell state representations}

Here we derive our representation of all encoded Bell states in terms of lower-encoding-level Bell states [Eqs. \eqref{eq:phys2block} and \eqref{eq:block2logic} in the main text],
\begin{align}
	\ket{\phi_{k,l}}^{(m)} &\cong \frac{1}{\sqrt{2^{m-1}}}\sum_{\vec{r} \in A_{l,m}} \bigotimes_{i=1}^m \ket{\phi_{k,r_i}},\label{eq:phys2blocka}\\
	\ket{\phi_{k,l}}^{(n,m)} &\cong \frac{1}{\sqrt{2^{n-1}}}\sum_{\vec{s} \in A_{k,n}} \bigotimes_{i=1}^n \ket{\phi_{s_i,l}}^{(m)},\label{eq:block2logica}
\end{align}
with $A_{l,m} = \Set{\vec{r}\in\{0,1\}^m | \sum_{i=1}^m r_i = l \Mod 2 }$.

Remember that QPC($n,m$) is constructed by $\ket{0}^{(m)} = \ket{0}^{\otimes m}$ [$\ket{1}^{(m)} = \ket{1}^{\otimes m}$] and $\ket{\pm}^{(n,m)} = \left[\ket{\pm}^{(m)}\right]^{\otimes n}$. We therefore obtain on the block level
\begin{align}
	\ket{\phi_{k,l}}^{(m)} &= \frac{1}{\sqrt{2}}\left[\ket{0}^{\otimes m} \ket{k}^{\otimes m} + (-1)^l \ket{1}^{\otimes m} \ket{1-k}^{\otimes m} \right]\nonumber\\
	& \cong \frac{1}{\sqrt{2}} \left[\ket{0,k}^{\otimes m} + (-1)^l \ket{1,1-k}^{\otimes m} \right]\nonumber\\
	& = \frac{1}{\sqrt{2^{m+1}}} \left[\left(\ket{\phi_{k,0}} + \ket{\phi_{k,1}}\right)^{\otimes m} + (-1)^l \left(\ket{\phi_{k,0}} - \ket{\phi_{k,1}}\right)^{\otimes m} \right]\nonumber\\
	&= \frac{1}{\sqrt{2^{m+1}}} \left[\sum_{\vec{r} \in \{0,1\}^m} \bigotimes_{i=1}^m \ket{\phi_{k,r_i}} + (-1)^l \sum_{\vec{r} \in \{0,1\}^m} \bigotimes_{i=1}^m (-1)^{r_i} \ket{\phi_{k,r_i}} \right]\nonumber\\
	&= \frac{1}{\sqrt{2^{m+1}}} \sum_{\vec{r} \in \{0,1\}^m} \left[1 + (-1)^{l + \sum_{j=1}^m r_j} \right] \bigotimes_{i=1}^m \ket{\phi_{k,r_i}}\nonumber\\
	&= \frac{1}{\sqrt{2^{m-1}}}\sum_{\vec{r} \in A_{l,m}} \bigotimes_{i=1}^m \ket{\phi_{k,r_i}}, \label{eq:phys2blockderivation}
\end{align}
where the symbol $\cong$ indicates the mode reordering mentioned in the main text. On the other hand, independent of the encoding, the Bell states can be written in the Pauli $X$-basis as
\begin{align*}
	\ket{\phi_{k,l}}^* &= \frac{1}{\sqrt{2}}\left[\ket{0,k}^* + (-1)^l \ket{1,1-k}^*\right]\\
	& = \frac{1}{2\sqrt{2}} \left[ \left(\ket{+}^*+\ket{-}^*\right)\left(\ket{+}^*+(-1)^k\ket{-}^*\right) + (-1)^l \left(\ket{+}^*-\ket{-}^*\right)\left(\ket{+}^*-(-1)^k\ket{-}^*\right) \right]\\
	& = \frac{1}{\sqrt{2}} \left[ \ket{(-1)^l +,+}^* + (-1)^k \ket{(-1)^l -,-}^* \right].
\end{align*}
On the logical level, we use this to obtain
\begin{align*}
	\ket{\phi_{k,l}}^{(n,m)} &= \frac{1}{\sqrt{2}} \left[ \left(\ket{(-1)^l+}^{(m)}\right)^{\otimes n}  \left(\ket{+}^{(m)}\right)^{\otimes n} +(-1)^k \left(\ket{(-1)^l-}^{(m)}\right)^{\otimes n}  \left(\ket{-}^{(m)}\right)^{\otimes n} \right]\\
	& \cong \frac{1}{\sqrt{2}} \left[ \left(\ket{(-1)^l+,+}^{(m)}\right)^{\otimes n} +(-1)^k \left(\ket{(-1)^l-,-}^{(m)}\right)^{\otimes n} \right]\\
	& = \frac{1}{\sqrt{2^{n+1}}} \left[\left(\ket{\phi_{0,l}}^{(m)} + \ket{\phi_{1,l}}^{(m)}\right)^{\otimes n} +(-1)^k \left(\ket{\phi_{0,l}}^{(m)} - \ket{\phi_{1,l}}^{(m)}\right)^{\otimes n} \right].
\end{align*}
At this point it becomes clear that, compared to the derivation on the block level \eqref{eq:phys2blockderivation}, the indices $k$ and $l$ have simply swapped their roles. This yields \eqref{eq:block2logica} immediately.

\subsection{Alternative derivation of the Bell measurement success probability}

The derivation of Eq. \eqref{eq:psucc} in the main text
\begin{align}
	p = \left[1-\left(1-\eta\right)^m\right]^n - \left[1-\left(1-\eta\right)^m - \tfrac{\eta^m}{2}\right]^n, \label{eq:psucca}
\end{align}
is based on the probability of loosing a photon on the physical level ($1-\eta$) and then calculating the chance of errors on the higher encoding levels with the help of the Bell state representations \eqref{eq:phys2block} and  \eqref{eq:block2logic}. We refer to this method of tracing the influence of the loss channel from bottom to top through the encoding levels as \textit{propagation of errors}. It is very versatile and is presented in much more detail in an extended treatment of the present results \cite{Ewert2016draft}, where it is also generalized to include multiple other error sources.

Here we present a different approach to deriving the Bell measurement success probability \eqref{eq:psucca}. It lacks the generality of the \textit{propagation of errors} and is not as easily expandable to include other error types than loss, but it might be more intuitive. In this approach the efficiency of an encoded Bell measurement is quantified by a set of values $p_\mu$, which are the probabilities of a successful Bell measurement if in total $\mu$ photons have been lost. We have already seen that a successful Bell measurement is possible as long as, with regards to the relevant logical qubit that was subject to loss in the communication channel, at least one photon is left in every block and at least one block is not corrupted by loss at all. This implies that the maximal number of photons that can be lost is $(n-1)(m-1)$. (This is much larger than the distance of the quantum parity code, which is $\min(n,m)-1$.)

In the main text we already calculated the value of $p_0$: $\mu=0$ corresponds to the ideal case of no losses, where we get $p_0=1-2^{-n}$. For $\mu \geq 1$ we get
\begin{align}
	p_\mu &= \frac{1}{\binom{nm}{\mu}} \sum_{i=0}^\mu\left(1-2^{-(n-i)}\right) \binom{n}{i}\underbrace{\sum_{\substack{j_1,...,j_i=1\\\sum\limits_{k=1}^i j_k =\mu}}^{m-1} \prod_{k=1}^i \binom{m}{j_k}}_{\mathcal{N}_{i,\mu,m}}. \label{eq:p_mu_one}
\end{align}
With the conventions $\mathcal{N}_{i,0,m} = \delta_{i,0}$ and $\mathcal{N}_{0,\mu,m} = \delta_{\mu,0}$ this also covers the case $\mu=0$. Although looking rather complicated at first and not being very convenient for numerical calculations, this formula can be understood quite nicely. The term $\mathcal{N}_{i,\mu,m}$ is the number of unique ways to distribute $\mu$ photon losses on exactly $i$ different blocks, such that in every one of these $i$ blocks at least one of the $m$ photons is not lost. The binomial coefficient $\binom{n}{i}$ represents the number of unique ways to pick $i$ of the $n$ blocks of the code. These are the $i$ corrupted blocks. The normalizing factor $\binom{n m}{\mu}$ is the number of unique ways to distribute $\mu$ photon losses on the $n m$ photons of the code. Ignoring for the moment the weight $\left(1-2^{-(n-i)}\right)$ the value $p_\mu$ is calculated simply by counting those loss configurations that allow for a successful Bell measurement. The weight $\left(1-2^{-(n-i)}\right)$ is the probability to identify the index $l$ in at least one of the remaining uncorrupted blocks (the corrupted blocks are useless for this, as discussed in the main text). The argument that this probability is given by $1-2^{-(n-i)}$ is the same as that for the derivation of $p_0$. In fact, the BM on the uncorrupted blocks can be seen as a BM on QPC($n-i,m$). A table with the values $p_{\mu}$ for various code sizes is given in section \ref{sec:table} of this supplemental material.

Next we want to show that the description of the Bell measurement efficiency with the set $p_\mu$ also yields the same value for the total BM success probability $p$. When all $n m$ photons are subject to the same loss channel $\eta$, the chance of loosing exactly $\mu$ photons is given by $\binom{n m}{\mu} \eta^{n m-\mu} (1-\eta)^{\mu}$. Therefore, the total BM success probability is
\begin{align}
	p' = \sum_{\mu=0}^{(n-1)(m-1)} p_{\mu}\binom{nm}{\mu} \eta^{nm-\mu}(1-\eta)^{\mu}.
\end{align}
To prove $p=p'$ we use the notation $[x^\mu]g(x)$, which is the coefficient of the monomial $x^\mu$ in the polynomial $g(x)$, for example $[x^\mu](1+x)^m = \binom{m}{\mu}$.
We get
\begin{align*}
	(1+x)^m & = \sum_{j=0}^m \binom{m}{j} x^j\\
	(1+x)^m - 1- x^m & = \sum_{j=1}^{m-1} \binom{m}{j} x^j\\
	[(1+x)^m - 1- x^m]^i & = \sum_{j_1,...,j_i=1}^{m-1} \left[\prod_{k=1}^i \binom{m}{j_k}\right] x^{j_1+...+j_k}.
\end{align*}
This yields
\begin{align*}
	\mathcal{N}_{i,\mu,m} & = \sum_{\substack{j_1,...,j_i=1\\\sum\limits_{k=1}^i j_k =\mu}}^{m-1} \prod_{k=1}^i \binom{m}{j_k} = [x^\mu] [(1+x)^m - 1- x^m]^i\\
	\Rightarrow \quad p_\mu &=  \frac{1}{\binom{nm}{\mu}} [x^\mu] \sum_{i=0}^\mu \binom{n}{i} \left[1-2^{-(n-i)}\right] [(1+x)^m - 1- x^m]^i\\
	&= \frac{1}{\binom{nm}{\mu}} [x^\mu] \left\{\left[(1+x)^m-x^m\right]^n - \left[(1+x)^m-x^m - \tfrac{1}{2} \right]^n \right\}\\
	\Rightarrow p' &= \sum_{\mu=0}^{(n-1)(m-1)} [x^\mu] \left\{\left[(1+x)^m-x^m\right]^n - \left[(1+x)^m-x^m - \tfrac{1}{2} \right]^n \right\} \left(\frac{1-\eta}{\eta}\right)^\mu \eta^{n m}\\
	&= \eta^{n m} \left\{\left[\frac{1}{\eta^m}-\frac{(1-\eta)^m}{\eta^m}\right]^n - \left[\frac{1}{\eta^m}-\frac{(1-\eta)^m}{\eta^m} - \frac{1}{2}\right]^n \right\}\\
	&= \left[1-\left(1-\eta\right)^m\right]^n - \left[1-\left(1-\eta\right)^m - \tfrac{\eta^m}{2}\right]^n = p.
\end{align*}

\subsection{Effect of faulty ancilla Bell states}
\label{subsec:faultyAnc}

In the main text we have assumed that the photon loss occurs only during the transmission and thus the probability of a (partially, i.e. $k=0$, $l=\ ?$) successful Bell measurement is given by $\eta=\exp(-L_0/L_{\rm att})$. However, possible imperfections in the ancillary Bell states should not be neglected. Here we assume the imperfections to be missing single photons within the QPC Bell state. Furthermore, we assume that the probability $(1-\eta_m)$ for a single photon of the QPC Bell state to be missing is the same for every photon of the state. The resulting mixed state can also be obtained by applying a photon loss error channel of strength $\eta_m$ to every mode of a perfect QPC Bell state. Since the composition of two photon loss channels yields another photon loss channel, the ``loss'' on the ancillary Bell state can be incorporated into the loss error channel of the transmission. In this context it should be noted that the two halves of the Bell state are treated differently: one is consumed immediately at the BM, whereas the other one is sent onward to the next repeater station. It is therefore useful to shift the point of view from the actual error channels to the success probabilities of a physical BM. A physical Bell measurement is (partially) successful if both photons are eventually measured. To this end, one photon must have been sent from the previous repeater station, i.e. it must not have been missing, and it must not be lost during the transmission. Additionally, the second photon which is part of the ancillary Bell state of the current repeater station must be present as well. We therefore get $\eta = \eta_m^2 \exp(-L_0/L_{\rm att})$. (At this point it is easy to also introduce lossy photon detectors.) Using this form of $\eta$ in the equations of the main text we obtain the total success rate in the case of ancilla states with missing photons. For example, we find for a communication distance of $L=\unit[1\,000]{km}$ and a chance of $1-\eta_m = 3\%$ to be missing a photon that the optimal code parameters $n=37$, $m=6$ and $L_0\approx\unit[2.09]{km}$ yield a success probability of $R t_0 = 73.46\%$.

Of course, missing single photons are not the only imperfections that can occur. However, with the above idea of dividing the errors on the encoded Bell state into two halves and incorporating them into the respective error channels leading to and from the repeater station, basically, every type of error that the quantum parity code can handle on the ``communication'' qubits can also be applied to the ancillary Bell states at the repeater stations. For more details, see the error rate analysis in \cite{Ewert2016draft}.

\newpage

\subsection{Generating QPC($n,m$)-states}

In order to generate both the encoded qubits and the ancillary encoded Bell states at the repeater stations we propose to employ a scheme based on coherent photon conversion (CPC), as it was  presented in Ref.~\cite{langford11}. Here a four-wave-mixing interaction,
\begin{align}
	H = \gamma a b^\dagger c^\dagger d + \gamma^* a^\dagger b c d^\dagger,
\end{align}
as realizable by a standard commercial, polarization-maintaining photonic crystal fiber (PCF), is pumped in one mode, e.g. that expressed by the annihilation operator $d$, with a bright classical beam to obtain effectively a three-wave-mixing Hamiltonian
\begin{align}
	\tilde{H} = \tilde{\gamma} a b^\dagger c^\dagger + \tilde{\gamma}^* a^\dagger b c
\end{align}
with a strong, tunable, nonlinear coupling $\tilde{\gamma}\propto \gamma E_d$ where $E_d$ is the (tunable) electric field amplitude of the pumping beam. The Hilbert space $\{\ket{1,0,0},\ket{0,1,1} \}$ is an eigenspace of this Hamiltonian $\tilde{H}$. Therefore, a state from this Hilbert space undergoes Rabi-like oscillations, when evolving under $\tilde{H}$. Especially, for an appropriate combination of coupling strength and interaction time, namely $\tfrac{\tilde{\gamma} t}{\hbar} = \tfrac{\pi}{2}$ this can be used as a photon doppler, because
\begin{align*}
	e^{-i\frac{\pi}{2} \tilde{H}} \ket{100} = \ket{011}.
\end{align*}
As a result, the two photons are now in modes $b$ and $c$. If desired, a transformation into two modes of frequency $\omega_a$ is possible with the use of two more CPC elements and by weakly pumping the remaining mode ($c$ or $b$). For more details, see Fig.~1b of Ref.~\cite{langford11}.

Since the vacuum passes the CPC element unchanged, the photon doppler setup can be used to transform qubit states into Bell- or GHZ-type states:
\begin{align}
	e^{-i\frac{\pi}{2} \tilde{H}} \left( \alpha \ket{0} + \beta \ket{1} \right) \otimes \ket{00} = \ket{0} \otimes \left( \alpha \ket{00} + \beta \ket{11} \right)
\end{align}
With the help of polarizing beam splitters it is also possible to build up a scheme that can do the same for polarization encoded qubits (see Fig.~2a) and b) in the Supplementary Material of Ref.~\cite{langford11}):
\begin{align*}
	\alpha\ket{H} + \beta \ket{V} \quad \adjustbox{raise={-2.5mm}}{\includegraphics[scale=0.7]{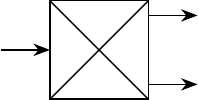}} \quad \alpha\ket{HH}+ \beta\ket{VV}
\end{align*}
When concatenating this photon doppling and using half wave plates oriented at $22.5^\circ$ to the optical axis (which realizes a Hadamard gate on polarization-encoded qubits), a scheme for generating arbitrary QPC($n,m$)-states is obtained:

\begin{figure}[H]
	\centering \includegraphics[scale=1]{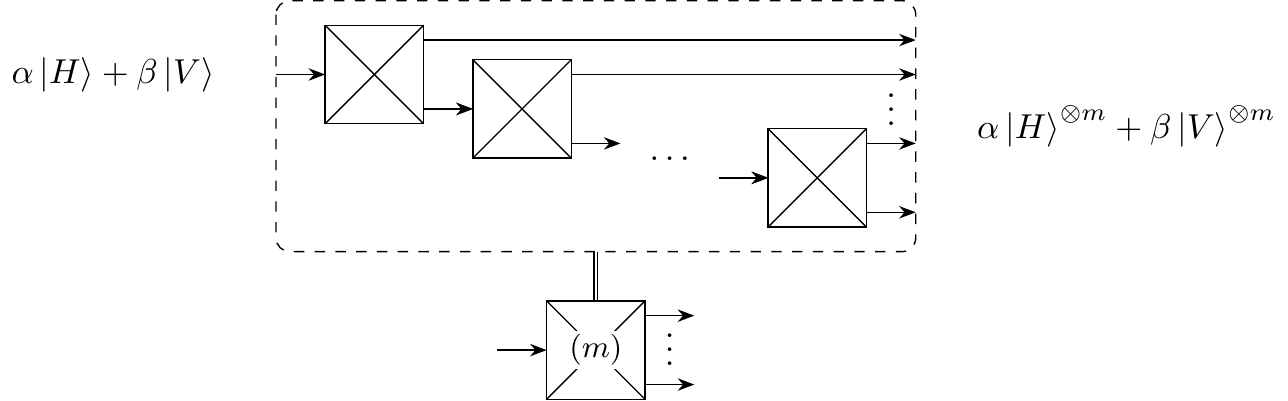}
	\caption{Concatenating $m-1$ photon dopplers gives a GHZ-type state with $m$ photons. \label{fig:m_doppler}}
\end{figure}
\begin{figure}[H]
	\centering \includegraphics[scale=1]{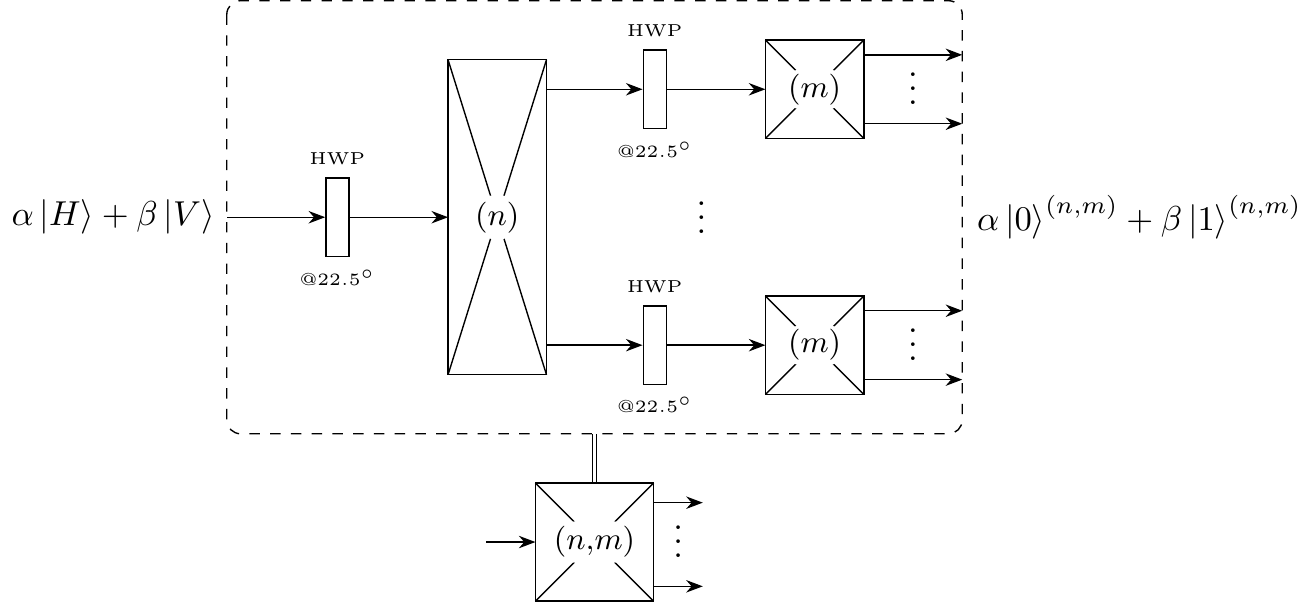}
	\caption{To generate an arbitrary QPC($n,m$) encoded state, an array of $n m -1$ photon dopplers and $n+1$ half-wave-plates, i.e. Hadamard gates, is used. \label{fig:QPC_Gen}}
\end{figure}
\begin{figure}[H]
	\centering \includegraphics[scale=1]{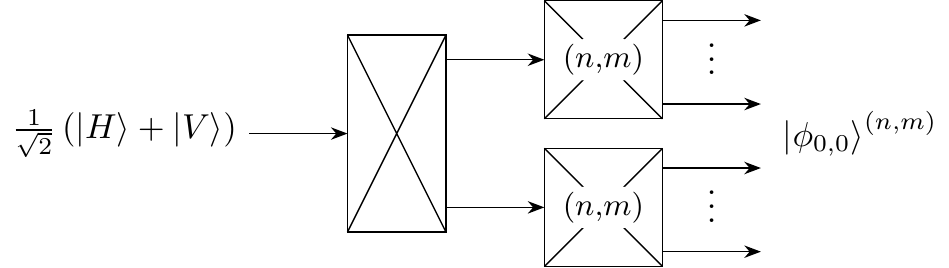}
	\caption{The ancillary Bell states at the repeater stations can be produced using two copies of the QPC($n,m$) generation scheme and one photon doppler to obtain the inital Bell state. This setup requires a total of $2 n m-1$ photon doppler modules. \label{fig:QPC_Bell_Gen}}
\end{figure}

A state generation scheme as depicted in Figs.~\ref{fig:m_doppler}~-~\ref{fig:QPC_Bell_Gen} gives exactly the (linear) cost scaling of $n m$ upon which the cost function $C$ as given in the main text relies, since for creating a QPC($n,m$)-encoded Bell state $2 n m-1$ copies of the photon doppler scheme are required.

It should be noted, however, that in this state generation scheme we depend on a strong nonlinear interaction, realizable with the techniques of coherent photon conversion. These nonlinearities could, in principle, also be used to obtain a unit-efficiency Bell measurement on the physical level. Employing these in our communication scheme instead of the standard linear optics BMs would allow the use of smaller quantum parity codes. However, the main point of our proposal is to present a highly effective and loss resistant Bell measurement on the quantum parity codes with the comparatively simple tools of linear optics.

Next, we take a look at the fault resistance of this state generation scheme. As was explained in Sec.~\ref{subsec:faultyAnc} of this Supplementary Material, all errors that occur on individual physical-level qubits can be incorporated into the error channel of the transmission protocol and are taken care of there. However, errors that occur during the state generation often lead to non-local errors in the resulting state. We analyze three stages of the state generation.

The first stage is generating the initial one-photon state $\ket{\psi} = \alpha\ket{H}+\beta\ket{V}$. Current photon sources cannot produce pure states of this form on demand and instead give a mixed state with a vacuum portion $\rho = \eta_s \ket{\psi}\bra{\psi} + (1-\eta_s) \ket{0}\bra{0}$. The vacuum state passes the CPC-elements unchanged and, as a result, the outcome of the state generation scheme can then be written as $\eta_s \ket{\psi}^{(n,m)}\bra{\psi}^{(n,m)} + (1-\eta_s) \ket{0}^{\otimes 2 n m}\bra{0}^{\otimes 2 n m}$. For the case that such a multimode-vacuum emerges in a repeater station during a time interval $t_0$ instead of a QPC Bell state, the logical BM will (heraldedly) fail at this station and thus the transmission does not succeed. The transmission rate when including the corresponding probabilistic element from the imperfect photon sources becomes $R = \eta_s^{L/L_0} p^{L/L_0}/t_0$. Due to this exponential scaling, the value for $\eta_s$ must be extremely close to one to still obtain acceptable communication rates. For example, for a transmission distance of $\unit[1\,000]{km}$ and a repeater spacing of $\unit[2]{km}$ the vacuum probability $1-\eta_s$ must be smaller than $0.0014$ to allow for repeater rates $R>0.5/t_0$. Unfortunately, current photon sources do not reach this near-deterministic regime, but are more commonly at values of about $\eta_s\approx0.5$. Yet, by using multiple heralded photon sources and a little feedforward (so-called multiplexing at the state preparation stage), it is possible to obtain a sufficient photon generation probability. For example, ten sources with $\eta_s=0.5$ yield at least one photon with probability $1-(1-\eta_s)^{10} \approx 0.9990$. Note that, in this case, it is not necessary to have $10$ instances of the nonlinear photon multiplying scheme presented in the above figures, because the feedforward operation takes place before that.

The second stage of the state generation scheme includes all steps to transform the single photon state into an $n$-photon state  ($2n$ for the Bell state generation). Every photon is then the seed for one of the blocks of the QPC state. Should any of these photons be lost, the entire corresponding block is missing in the final state. This will lead to a (heralded) failure of the entire BM. Therefore the probability to lose any photons in this stage must be reduced as much as possible. Luckily, both the linear and the non-linear parts of the photon doppling can be performed near-deterministically \cite{langford11}. Should, nevertheless, the obtained success probability for the second stage of the state generation scheme be too small, multiplexing this stage as well would still be an option. Only very few instances would be required at every repeater station, as the base probability is already quite high. The important task of heralding the failure to initiate the feed-forward operation can be performed by generating an $n+1$-photon GHZ state instead of the $n$-photon GHZ state and measuring the additional photon in the $X$-basis. Due to the simple concatenation structure of the GHZ-state generation (see Fig.~\ref{fig:m_doppler}), the last photon is only present if all other $n$ photons are as well. The possible (heralded) phase-flip induced by the $X$-basis measurement can easily be corrected with the help of a $Z$-gate on any of the remaining modes. Using, for example, three multiplying instances (Fig.~\ref{fig:m_doppler}) with a total success rate of $0.9$ each, together with 10 photon sources of efficiency $\eta_s=0.5$ per instance, we obtain a probability of $\approx 0.9990$ for a successful generation of the desired GHZ state.

In the third stage of the generation scheme, the individual blocks are built up. Should a photon be lost in this stage, the resulting block can still be used for the identification of the Bell-state index $k$ in this block, as long as at least one physical BM succeeds. Only its ability to identify the index $l$ is gone. The remaining block is thus still useful, even though with the loss of one photon no more extra photons will be added to the state due to the linear concatenation scheme in Fig.~\ref{fig:m_doppler}. The number of lost photons can be further decreased by using a more tree-like concatenation structure. Then only those modes of the branches starting at the point of loss will be empty instead of all later modes. This decreased number of photon loss increases the chance for at least one physical BM to be successful. We expect the effect of losses in this stage on the transmission rate to be quite small, as long as the chance to lose photons particularly early within the third stage is not too high. 

Our communication scheme is not at all restricted to be based on the nonlinear state generation scheme as described above. We may also consider employing exclusively linear optical methods which then includes the generation of the encoded states at every repeater station. Such an approach is similar to the all-optical repeater schemes proposed in \cite{ATL,pant2016rate}. We found that a toolbox for state preparation that contains single-photon sources and detectors, beam splitters, and feedforward operations \cite{ATL,pant2016rate} is also sufficient to create at every repeater station the QPC-encoded states that are needed in our scheme \cite{Ewert2016draft}. These states can be obtained, again similar to \cite{ATL,pant2016rate}, by first creating so-called tree-cluster states (which themselves are obtainable from three-photon GHZ states, probabilistically generated from the initial single-photon states, using Bell measurements) and connecting these through Bell measurements. For a QPC as big as $(24,5)$, which is similar to our optimal code in the presence of only transmission losses, an average number of $10^{6}$ photons is needed at every station in order to ensure that every time unit $t_0$ at least one copy of an encoded state is available \cite{Ewert2016draft}. Interestingly, this number is of the same order of magnitude as that derived in \cite{pant2016rate} where an improved state generation scheme was proposed compared to \cite{ATL} (for our estimation, we assumed perfect single-photon sources and advanced $3/4$-Bell measurements \cite{Ewert}, where indeed performing improved Bell measurements helps a lot for the state generations while it is unnecessary for the actual repeater protocol, see main text and \cite{Ewert2016draft} -- in contrast to \cite{pant2016rate} where it helps for both). This may be owing to the fact that the encoded states from the two schemes and especially their basic tree-type resource states resemble each other from a conceptual point of view.

Note that in either case, a huge amount of feedforward is needed in order to achieve the multiplexing required at every stage of the state generation (in this respect, the multiplexing as described above in front of the CPC setup is much cheaper in terms of feedforward). Moreover, although the overhead of single photons at every station may be constant with distance for reasonably long total distances, it is still a rather demanding overhead from a realistic, experimental point of view. Nonetheless, considering this kind of feedforward-based linear-optical state generation also for our scheme, we emphasize that only in our scheme there is still no need for any ``online'' feedforward operations which, in contrast, are a crucial ingredient in \cite{ATL,pant2016rate}. Similarly, only in our scheme, there is no need to distribute non-locally the encoded, massively entangled (cluster) states between repeater stations; our encoded Bell states are created and consumed locally and only the encoded input qubit must be sent over the transmission channel. Thus, there is a significant conceptual difference between our forward-error-corrected, static linear-optics scheme and those non-local-entanglement- and feedforward-based linear optics schemes of \cite{ATL,pant2016rate}. 

Other linear optical methods, in particular, those based on percolation theory [PRL 115, 020502 (2015)],[PRA 91, 042301 (2015)] could, in principle, be used to circumvent the need for feedforward to some extent in any of these schemes. Such an approach, of course, would be most compatible with our static linear-optics protocol.
However, although the method for the ``offline'' creation of QPC-Bell states with the nonlinear tools of CPC as described above leaves the realm of linear optics, it still has the appealing property (in addition to its linear cost in terms of single photons) that it is all-optical (no atomic qubits are required for processing) and as well static (no complicated feedforward techniques are absolutely necessary). That is why, in the present proposal, we take it as the method of our choice.

\subsection{Bell measurement efficiencies}
\label{sec:table}
In the following table the success probabilities $p_\mu$ of the BM according to Eq. \eqref{eq:p_mu_one} are given for various QPC$(n,m)$ encodings.
\newpage
\begin{turnpage}
\setlength{\tabcolsep}{2mm}
\setlength{\extrarowheight}{2mm}
\begin{table}
\caption{Success probabilities $p_\mu$ in $\%$ of the BM given the total number of photons lost $\mu$ on one of the logical qubits.}
\begin{tabular}{c|lllllllllllllllllll}

($n,m$) & 0 & 1 & 2 & 3 & 4 & 5 & 6 & 7 & 8 & 9 & 10 & 11 & 12 & 13 & 14 & 15 & 16 & 17 & 18  \\
\hline
(1,1) & 50 &&&&& &&&&& &&&&& &&& \\
(1,2) & 50 &&&&& &&&&& &&&&& &&& \\
(2,1) & 75	&&&&& &&&&& &&&&& &&& \\
(2,2) & 75 & 50	&&&& &&&&& &&&&& &&& \\
(2,3) & 75 & 50 & 20 &&& &&&&& &&&&& &&& \\
(3,2) & 87.5 & 75 & 40 &&& &&&&& &&&&& &&& \\
(3,3) & 87.5 & 75 & 56.25 & 32.14 & 10.71 & &&&&& &&&&& &&& \\
(3,4) & 87.5 & 75 & 56.82 & 36.82 & 20.61 & \hphantom{0}9.09 & \hphantom{0}2.60 &&&& &&&&& &&& \\
(4,3) & 93.75 & 87.5 & 77.27 & 61.36 & 40.91 & 20.45 & \hphantom{0}5.84 &&&& &&&&& &&& \\
(3,5) & 87.5 & 75 & 57.14 & 37.91 & 22.80 & 12.49 & \hphantom{0}5.99 & \hphantom{0}2.33 & \hphantom{0}0.58  && &&&&& &&& \\
(5,3) & 96.88 & 93.75 & 88.39 & 79.12 & 65.11 & 47.20 & 28.32 & 12.59 & \hphantom{0}3.15 && &&&&& &&& \\
(4,4) & 93.75 & 87.5 & 77.50 & 63.93 & 48.46 & 33.52 & 20.68 & 10.91 & \hphantom{0}4.48 & \hphantom{0}1.12 & &&&&& &&& \\
(4,5) & 93.75 & 87.5 & 77.63 & 64.47 & 49.90 & 36.28 & 24.90 & 16.06 & \hphantom{0}9.62 & \hphantom{0}5.21 & \hphantom{0}2.44 & \hphantom{0}0.89 & \hphantom{0}0.20 &&& &&& \\
(5,4) & 96.88 & 93.75 & 88.49 & 80.59 & 70.07 & 57.66 & 44.58 & 32.04 & 21.05 & 12.29 & \hphantom{0}6.06 & \hphantom{0}2.29 & \hphantom{0}0.51 &&& &&& \\
(5,5) & 96.88 & 93.75 & 88.54 & 80.84 & 70.84 & 59.41 & 47.75 & 36.85 & 27.30 & 19.35 & 13.04 & \hphantom{0}8.26 & \hphantom{0}4.83 & \hphantom{0}2.52 & \hphantom{0}1.12 & \hphantom{0}0.38 & \hphantom{0}0.08 && \\
(7,4) & 99.22 & 98.44 & 97.05 & 94.75 & 91.20 & 86.10 & 79.31 & 70.94 & 61.32 & 50.96 & 40.49 & 30.51 & 21.58 & 14.13 & \hphantom{0}8.38 & \hphantom{0}4.35 & \hphantom{0}1.87 & \hphantom{0}0.60 & \hphantom{0}0.11 \\
(3,10) & 87.5 & 75 & 57.76 & 39.90 & 25.94 & 16.19 & \hphantom{0}9.77 & \hphantom{0}5.71 & \hphantom{0}3.23 & \hphantom{0}1.76 & \hphantom{0}0.92 & \
\hphantom{0}0.46 & \hphantom{0}0.22 & \hphantom{0}0.10 & \hphantom{0}0.04 & \hphantom{0}0.01 & \hphantom{0}0.00 & \hphantom{0}0.00 & \hphantom{0}0.00\\
(6,5) & 98.44 & 96.88 & 94.18 & 89.95 & 83.92 & 76.15 & 67.05 & 57.26 & 47.45 & 38.15 &
29.73 & 22.40 & 16.26 & 11.30 & \hphantom{0}7.44 & \hphantom{0}4.58 & \hphantom{0}2.58 & \hphantom{0}1.29 & \hphantom{0}0.54\\
(10,3) & 99.90 & 99.80 & 99.62 & 99.05 & 97.76 & 95.39 & 91.59 & 86.01 & 78.45 & 68.91 &
57.70 & 45.47 & 33.22 & 22.05 & 12.95 & \hphantom{0}6.47 & \hphantom{0}2.59 & \hphantom{0}0.74 & \hphantom{0}0.11\\
(10,4) & 99.90 & 99.8 & 99.62 & 99.31 & 98.76 & 97.86 & 96.46 & 94.37 & 91.43 & 87.51 &
82.52 & 76.47 & 69.45 & 61.65 & 53.33 & 44.81 & 36.42 & 28.51 & 21.37\\
(12,4) & 99.98 & 99.95 & 99.91 & 99.82 & 99.67 & 99.41 & 98.96 & 98.25 & 97.17 & 95.59 &
93.41 & 90.52 & 86.82 & 82.27 & 76.88 & 70.70 & 63.86 & 56.53 & 48.92\\
(15,5) & 100.0 & 99.99 & 99.99 & 99.98 & 99.96 & 99.93 & 99.87 & 99.78 & 99.64 & 99.43 &
99.11 & 98.65 & 98.02 & 97.17 & 96.06 & 94.65 & 92.90 & 90.79 & 88.30\\
(23,5) & 100.0 & 100.0 & 100.0 & 100.0 & 100.0 & 100.0 & 100.0 & 100.0 & 100.0 & 99.99 & 99.99 & 99.98 & 99.97 & 99.96 & 99.93 & 99.90 & 99.85 & 99.78 & 99.68\\
(30,6) &100.0&100.0&100.0&100.0&100.0&100.0&100.0&100.0&100.0&100.0&100.0&100.0&100.0&100.0&100.0&100.0&100.0&100.0&100.0
\end{tabular}
\label{table:probsa}
\end{table}
\end{turnpage}
\clearpage
\global\pdfpageattr\expandafter{\the\pdfpageattr/Rotate 90}
\clearpage
\global\pdfpageattr\expandafter{\the\pdfpageattr/Rotate 0}
\end{document}